\documentclass[twocolumn,tighten]{aastex63}
\usepackage{graphicx,amsmath,natbib}

\shorttitle{Barium in M15}
\shortauthors{Kirby et al.}
\newcommand{\bahlo}{-1.91}
\newcommand{\bahloerr}{0.15}
\newcommand{\bahmed}{-2.13}
\newcommand{\bahmederr}{0.03}
\newcommand{\bahhi}{-2.23}
\newcommand{\bahhierr}{0.02}
\newcommand{\nba}{96}
\newcommand{\ngood}{66}
\newcommand{\nbad}{15}
\newcommand{\nul}{15}

\begin{document}

\title{The Stars in M15 Were Born with the $r$-process\footnote{The
    data presented herein were obtained at the W.~M.~Keck Observatory,
    which is operated as a scientific partnership among the California
    Institute of Technology, the University of California and the
    National Aeronautics and Space Administration. The Observatory was
    made possible by the generous financial support of the W.~M.~Keck
    Foundation.}}

\correspondingauthor{Evan N. Kirby}
\email{enk@astro.caltech.edu}

\author[0000-0001-6196-5162]{Evan N. Kirby}
\affiliation{California Institute of Technology, 1200 E.\ California Blvd., MC 249-17, Pasadena, CA 91125, USA}

\author[0000-0002-9256-6735]{Gina Duggan}
\affiliation{California Institute of Technology, 1200 E.\ California Blvd., MC 249-17, Pasadena, CA 91125, USA}

\author[0000-0003-2558-3102]{Enrico Ramirez-Ruiz}
\affiliation{Department of Astronomy and Astrophysics, University of California, Santa Cruz, CA 95064, USA}
\affiliation{Niels Bohr Institute, University of Copenhagen, Blegdamsvej 17, DK-2100 Copenhagen, Denmark}

\author[0000-0002-9946-4635]{Phillip Macias}
\affiliation{Department of Astronomy and Astrophysics, University of California, Santa Cruz, CA 95064, USA}


\begin{abstract}
High-resolution spectroscopy of stars on the red giant branch (RGB) of
the globular cluster M15 has revealed a large ($\sim 1$~dex)
dispersion in the abundances of $r$-process elements, like Ba and Eu.
Neutron star mergers (NSMs) have been proposed as a major source of
the $r$-process.  However, most NSM models predict a delay time longer
than the timescale for cluster formation.  One possibility is that a
NSM polluted the surfaces of stars in M15 long after the cluster
finished forming.  In this case, the abundances of the polluting
elements would decrease in the first dredge-up as stars turn on to the
RGB\@.  We present Keck/DEIMOS abundances of Ba in \ngood\ stars along
the entire RGB and the top of the main sequence.  The Ba abundances
have no trend with stellar luminosity (evolutionary phase).
Therefore, the stars were born with the Ba they have today, and Ba did
not originate in a source with a delay time longer than the timescale
for cluster formation. In particular, if the source of Ba was a
neutron star merger, it would have had a very short delay
time. Alternatively, if Ba enrichment took place before the formation
of the cluster, an inhomogeneity of a factor of 30 in Ba abundance
needs to be able to persist over the length scale of the gas cloud
that formed M15, which is unlikely.
\end{abstract}


\section{Introduction}
\label{sec:intro}

Elements beyond iron in the periodic table are made primarily via
neutron capture, which can happen either slowly ($s$-process) or
rapidly ($r$-process).  While the $s$-process is known to occur
primarily in asymptotic giant branch stars \citep{tru77,kar14}, there
are still multiple candidate sites for the $r$-process
\citep{lat74,qia07}.  The gravitational wave-based discovery and
subsequent electromagnetic observation of a kilonova in 2017
definitively showed that the $r$-process does occur in neutron star
mergers \citep[NSMs, e.g.,][]{kas17}.  However, it is far from clear
that NSMs are the sole site of the $r$-process \citep{hot18,cot19}.

The strongest argument that the $r$-process is also created in other
events is that NSMs are expected to have long delay times
\citep[$\gtrsim 10^8$~yr,][]{kal01}.  For example, it is difficult to
explain the early appearance of the $r$-process in the Milky Way (MW)
halo when it is assumed that metals are instantaneously mixed
\citep{van19}.  However, a more realistic treatment of mixing has
resulted in differing conclusions on whether NSMs can explain the
[Eu/Fe] scatter at low metallicities in the halo \citep{nai18,hay19}.

It is possible that the $r$-process source at late times (or high
metallicity) is different from the source at early times (or low
metallicity).  At low metallicities in the MW halo \citep{mac18} and
ultra-faint dwarf galaxies \citep{ji16b}, the source seems to be rare
and prolific, producing at least $10^{-3}~M_{\sun}$ of $r$-process
material per event.  On the other hand, the $r$-process element Eu in
the MW disk appears to be created in lockstep with Mg
\citep[e.g.,][]{ish13}, which is nearly instantaneously recycled from
core collapse supernovae (CCSNe).  Therefore, it seems that the
$r$-process source is not delayed at high metallicity or late times.
\citet{dug18} showed that the $r$-process component of Ba is delayed
relative to CCSNe at ${\rm [Fe/H]} < -1.6$---but perhaps not at higher
metallicities---in the Sculptor dwarf galaxy.  However, \citet{sku19}
argued that Eu in Sculptor shows no delay relative to Mg, although
their data was limited in the metallicity range where
\citeauthor{dug18}\ showed that $r$-process Ba is delayed.

There are two possibilities for synthesizing the $r$-process with a
short delay.  First, it is possible that there is a prompt population
of NSMs.  For example, the dynamics of dense star clusters can shorten
the delay time compared to a NSM formed in the field \citep{ram15}.
Second, there could be another major source of the $r$-process.
Although the high-entropy wind surrounding the proto-neutron star of a
CCSN was initially thought to be a promising site for $r$-process
production \citep{mey92}, most CCSN simulations fail to achieve the
conditions required for the $r$-process \citep[e.g.,][]{qia96}.
Alternatives include jet-driven, magnetorotational CCSNe \citep{nis15}
or the accretion disks around the supernovae that result from rapidly
rotating massive stars \citep[collapsars,][]{sie19}.  One issue with
forming the $r$-process from massive stars is that observed stellar
abundance distributions of metal-poor MW stars disfavor an $r$-process
site that also produces iron \citep{mac19}, which CCSNe are expected
to produce.  The current state of the field is that observations
indicate that some $r$-process material comes from prompt sources, but
it is difficult to produce the $r$-process in theoretical simulations
of prompt sources.

Globular clusters (GCs) might be able to help show a fuller range of
$r$-process production sites.  GCs are complex sites of star
formation.  Almost all GCs show multiple chemical populations
\citep{gra12}, but no theory proposed so far can explain all of the
nuances of the observed chemical abundance patterns \citep{bas18}.
The multiple populations in GCs are most evident in light elements,
like O and Na, but a small number of GCs possibly shows variations in
neutron-capture elements \citep{roe11}.  However, the evidence of a
dispersion in some clusters has been called into question
\citep{coh11b}.  The only incontrovertible example of this phenomenon
is M15, which shows $\sim 1$~dex scatter in Ba, Eu, and other heavy
elements \citep{sne97,sne00a,sob11,wor13}.  The neutron-capture
abundance pattern in nearly all GCs, including M15, is dominated by
the $r$-process \citep[e.g.,][]{sne00a}.\footnote{The major exception
  is Omega Centauri, which is either a globular cluster or the nucleus
  of an accreted dwarf galaxy \citep{maj00}.  It has a large
  dispersion in [Ba/H], but the Ba seems to have been created in the
  $s$-process rather than the $r$-process \citep{smi00}.}  Therefore,
some phenomenon must be able to pollute M15 inhomogeneously with the
$r$-process.  The inhomogeneity arises from a different source from
the light elements because there is no correlation between the
abundance variations in the light and neutron-capture elements in M15
or any other GC \citep{roe11}.

In this letter, we investigate the possibility that the $r$-process in
M15 was created by a NSM\@.  Specifically, we consider a NSM with a
``standard'' delay time ($>10^8$~yr), much longer than the timescale
of the formation of the cluster ($\sim 10^7$~yr).  In this scenario,
after the ejecta of the NSM sweeps up enough mass to cool
\citep{mon16}, it would pollute already-formed stars via Bondi
accretion.  This scenario explains the star-to-star scatter in
neutron-capture abundances because stars nearest to the NSM would have
received the highest degree of pollution \citep[see][]{tsu14}.  The
amount of $r$-process material in M15 is about the amount that a
single NSM is expected to generate, which allows this scenario to be
viable as long as most of the material is retained in the cluster.
From the observed Eu abundances \citep{wor13}, we estimate that the
stars in M15 contain about $8 \times 10^{-6}~M_{\sun}$ of Eu, compared
to $(3-15) \times 10^{-6}~M_{\sun}$ generated by a single NSM
\citep[e.g., GW170817,][]{cot18}.  The hypothesis predicts depletion
of neutron-capture elements as stars ascend the red giant branch
(RGB)\@.  We test this prediction with measurements of Ba abundances
in M15 from the main sequence to the tip of the RGB\@.


\section{Observations and Abundance Measurements}
\label{sec:obs}

We observed a single slitmask with DEIMOS on the Keck~II telescope on
2017 Sep 15.  The slitmask, called \texttt{7078l1}, was the same as
that observed by \citet{kir16}.  We used $BVRI$ photometry from
\citet{ste94}.  We chose targets from the RGB and main sequence
turn-off (MSTO).  The selection was performed by drawing a polygon
around the locus of stars in the color--magnitude diagram (CMD)\@.
The width of the polygon was about 0.7~mag in $B-V$ color, which is
wide enough to include effectively 100\% of candidate member stars.

\begin{figure}
\centering
\includegraphics[width=\linewidth]{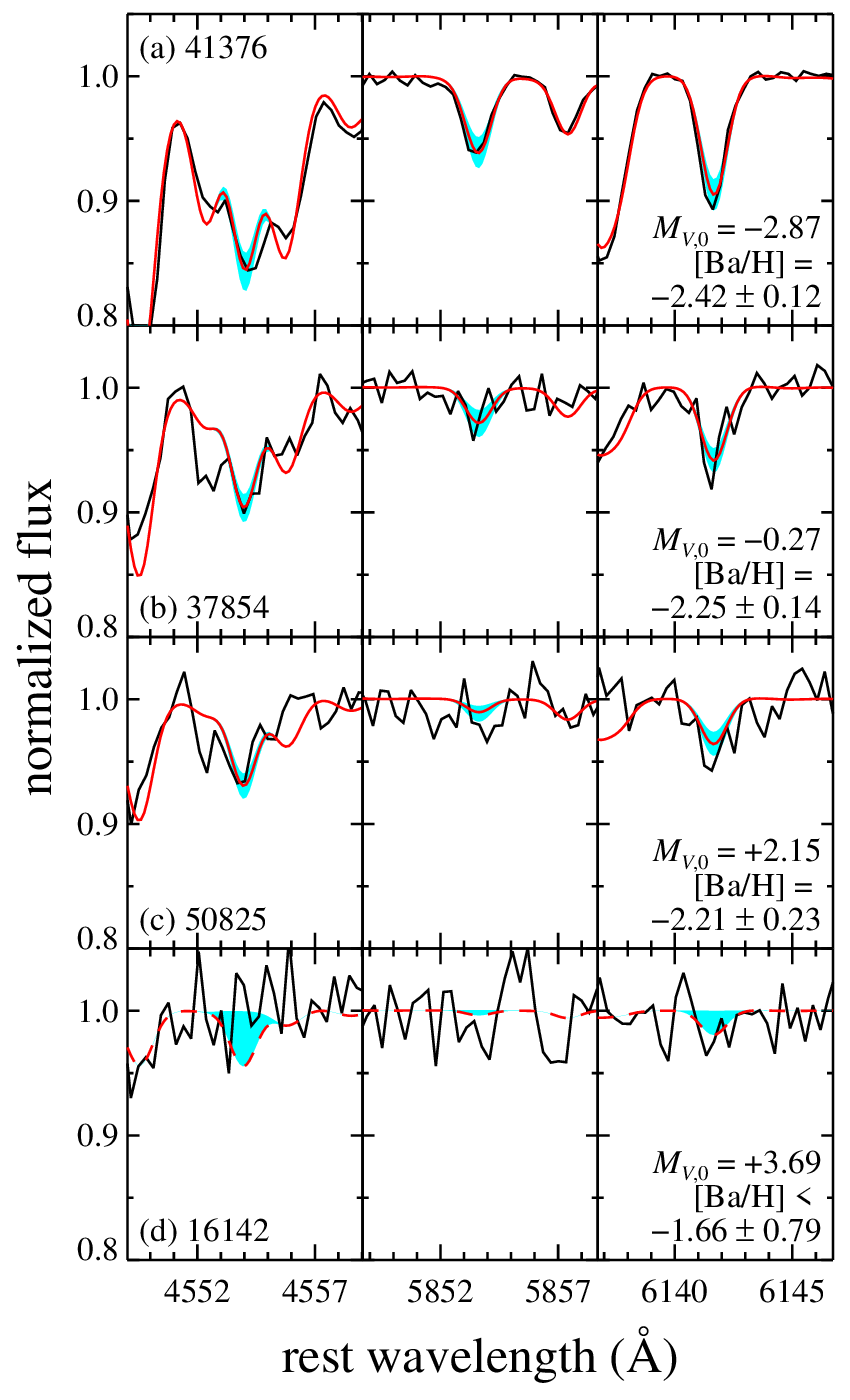}
\caption{DEIMOS spectra for stars at various evolutionary stages: (a)
  tip of the RGB, (b) near the red clump, (c) near the end of the FDU,
  and (d) just below the MSTO\@.  Each panel lists the object name,
  extinction-corrected absolute $V$ magnitude, and measured Ba
  abundance.  Only small regions around three Ba lines are shown.  The
  full spectral range is much larger, and the analysis uses two other
  Ba lines not shown here.  The best-fitting synthetic spectra are
  shown in red, and the cyan regions show the response to a change in
  [Ba/H] of $\pm 0.3$~dex.  The bottom panel shows an example of a
  90\% C.L.\ upper limit on Ba abundance, where the cyan region shows
  the response of the spectrum from changing the Ba abundance from the
  upper limit to zero.\label{fig:spectra}}
\end{figure}

We used the 1200B diffraction grating at a central wavelength of
5500~\AA\@.  The approximate spectral range was 3900--4500~\AA\ with a
resolution of $\Delta\lambda = 1.1$~\AA\@.  We obtained 13 exposures
of 20 minutes each for a total exposure time of 4.3 hours.  The seeing
was 0.6'', and the transparency was good.  We reduced the spectra with
\texttt{spec2d} \citep{coo12,new13}, including our own modifications
for the wavelength solution for the 1200B grating
\citep[see][]{del19}.

\citet{kir16} previously measured radial velocities, effective
temperatures ($T_{\rm eff}$), and metallicities ([Fe/H]) for the stars
on this slitmask using DEIMOS's 1200G diffraction grating.  We used
their membership determination, which enforced that members have
radial velocities and metallicities within 3 standard deviations of
their respective cluster means.  We discarded the spectra of known
non-members.

We measured Ba abundances using the procedure of \citet{dug18}.  They
constructed a grid of synthetic spectra over a range of $T_{\rm eff}$,
surface gravity ($\log g$), [Fe/H], and Ba abundance.  The spectra
were computed with MOOG \citep{sne12}.  The grid is searched for the
Ba abundance that minimizes $\chi^2$ between the grid and the observed
spectrum.  Some parameters ($T_{\rm eff}$ and $\log g$) were fixed at
their previously determined values \citep{kir16}.  However, we
enforced that [Fe/H] and [$\alpha$/Fe] were the same for each star.
In other words, we assumed that the intrinsic dispersion of these
abundances in the cluster is essentially zero.  The values were fixed
at the median of the values determined by \citet{kir16} for the stars
in our present sample (${\rm [Fe/H]} = -2.46$ and ${\rm [\alpha/Fe]} =
0.30$).\footnote{We also measured Ba abundances using [Fe/H] and
  [$\alpha$/Fe] determined for each star.  The mean Ba abundance
  decreased by 0.01~dex, and the standard deviation between the two
  methods was 0.08~dex.  We conclude that both methods yield
  essentially identical results.}

The uncertainties on the Ba abundances are a combination of random
uncertainty, which results from spectral noise, and systematic error,
which results from a variety of sources, including imperfections in
the spectral models.  The systematic error of 0.1~dex was determined
by \citet{dug18}.  It is added in quadrature with the random
uncertainty.  As a result, 0.1~dex is the minimum error that we quote.

In some cases, we could not measure Ba abundances with uncertainties
less than 0.5~dex.  We treat these cases as upper limits.  We quote
the 90\%~C.L.\ upper limit by finding the Ba abundance that bounds
90\% of the probability distribution.  In some other cases, we were
not even able to estimate an upper limit, due to excessive noise or
anomalies in the spectra that affected the Ba lines.
Table~\ref{tab:ba} gives the Ba abundances for all the M15 member
stars in our sample.  The table includes Ba measurements, upper
limits, and non-measurements.  The sample includes \nba\ stars,
comprised of \ngood\ Ba abundance measurements, \nul\ upper limits,
and \nbad\ stars for which we were not able to measure Ba abundance or
estimate an upper limit.

We used the local thermodynamic equilibrium (LTE) approximation in the
spectral models and the atmospheres used to generate them.  Some lines
of Ba are subject to non-LTE corrections.  \citet{dug18} investigated
the magnitude of the corrections \citep[based on the work of][]{and09}
for 12 stars observed with DEIMOS over the ranges $3747~{\rm K} <
T_{\rm eff} < 5075~{\rm K}$ and $0.21 < \log g < 2.00$.  The
corrections to [Ba/Fe] ranged from $-0.16$ to $+0.15$.  The more
recent work of \citet{eit19} found typical corrections of $-0.2$ for
red giants.  While these corrections would be important for some
applications, we use our sample to look for changes in [Ba/Fe] greater
than 1~dex (see Section~\ref{sec:mixing}).  Therefore, our main
conclusion is not affected by our approximation of LTE\@.

Figure~\ref{fig:spectra} shows the DEIMOS spectra of four stars at
different phases of evolution, from the main sequence to the tip of
the RGB\@.  The stars in the figure span a range of 6.6~mag in the $V$
band.  The corresponding range of spectral quality is apparent.  The
figure demonstrates how well the synthetic spectra fit the observed
spectra.  It also shows the response of the spectra to a change in Ba
abundance of $\pm 0.3$~dex.  The individual Ba absorption lines in the
fainter stars are at the edge of detectability.  However, the spectral
synthesis technique uses all of the information from five Ba lines
(two of which are not shown in Figure~\ref{fig:spectra})
simultaneously.  Thus, detections of individual lines are not a
requirement for measuring abundances.


\section{Mixing on the RGB}
\label{sec:mixing}

Low-mass stars, such as those in M15, experience mixing events as they
evolve off the main sequence \citep[see the review by][]{kar14}.  The
first mixing episode is the first dredge-up (FDU), which occurs after
the core is exhausted of hydrogen.  The resulting contraction of the
core drives the star to expand and its convective envelope to deepen.
The FDU brings products of hydrogen burning (e.g., ${}^{13}$C) to the
surface.  It also submerges and dilutes species that are only present
on the stellar surface.  The quintessential example of dilution at the
FDU is ${}^7$Li.  Because ${}^7$Li burns at a low temperature, it
exists only in the outer layers of the star.  The FDU dilutes the
surface abundance as Li-poor material is dredged up to the surface.

The second mixing episode occurs at the luminosity function bump in
the RGB\@.  The convective envelope retreats in mass coordinate when
the star is about halfway up the RGB\@.  The retreating envelope
leaves behind a discontinuity in mean molecular weight.  As the
hydrogen burning shell expands in mass coordinate, it eventually
crosses this discontinuity, causing a brief pause in the star's ascent
up the RGB\@.  It is here that ``extra mixing''---most likely
thermohaline mixing \citep{cha07}---has been observed in several
species, including C and Li \citep{gra00}.  The C and Li abundances
drop because extra mixing connects the convective envelope to
temperatures sufficient to burn ${}^{12}$C and ${}^{7}$Li through
proton capture.  This extra mixing is not expected to affect the
abundance of Ba.

If a NSM polluted the surfaces of stars in M15 long after they formed,
then the surface compositions of the stars would be different from
their centers.  Mixing at the FDU would dilute the $r$-process species
that originated in the NSM\@.  Therefore, we search for dilution
signatures of Ba in M15 at the stellar luminosity that corresponds to
the FDU\@.

To further quantify the observational signature of external pollution,
we used Modules for Experimentation in Stellar Astrophysics
\citep[MESA,][]{pax11} to simulate the dilution at the FDU of an
$r$-process pollution event.  We simulated the post-main sequence
evolution of a $0.8~M_{\sun}$ star with ${\rm [Fe/H]} = -2$, roughly
corresponding to stars at the MSTO in M15.  In the simulation, some
$r$-process material polluted the surface of the star when it was on
the main sequence.  Convection mixed this material throughout the
convective envelope but no deeper.  When the star turned off the main
sequence, the FDU depleted the surface abundance of ${}^{153}$Eu by a
factor of $\sim 30$, or 1.5~dex.  If this scenario is occurring in
M15, then we expect to see such a depletion in the abundances of all
$r$-process elements, including our Ba measurements.


\section{Results}
\label{sec:results}

\begin{figure*}
\centering
\includegraphics[width=0.97\linewidth]{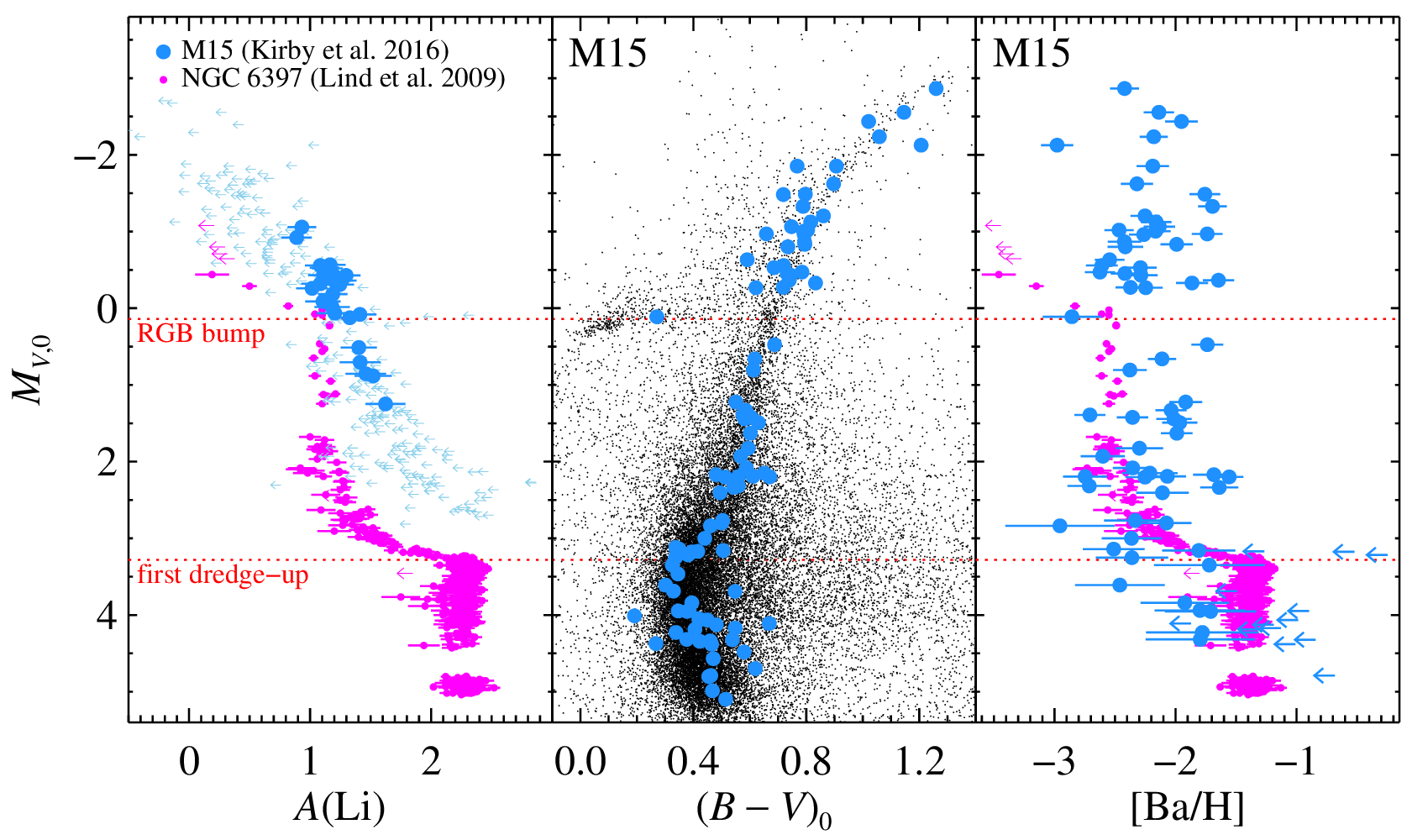}
\caption{{\it Left:} As stars evolve up the RGB, Li is diluted.
  Measurements in M15 from DEIMOS \citep{kir16} are shown as large
  blue points and light blue upper limits.  Higher quality
  measurements in NGC~6397 \citep[small magenta points and upper
    limits,][]{lin09} are also shown as a clearer demonstration of the
  dilution.  {\it Center:} The color--magnitude diagram of M15
  \citep[small points,][]{ste94}.  Stars with Ba abundance
  measurements from DEIMOS are shown as large blue points.  {\it
    Right:} DEIMOS measurements of [Ba/H] for stars at a variety of
  evolutionary phases in M15 (blue) including some upper limits
  (large, leftward-pointing blue arrows).  The Li abundances in
  NGC~6397, shifted by an arbitrary constant, are shown in magenta to
  illustrate the expected dilution if Ba were present only on the
  stellar surface.  The blue points in the center and right panels
  represent the same stars.  Red dotted lines indicate the absolute
  magnitudes of the FDU and RGB bump. \label{fig:cmdbali}}
\end{figure*}

Figure~\ref{fig:cmdbali} summarizes our abundance measurements.  The
central panel shows the positions of our spectroscopic targets in the
CMD in terms of absolute magnitude, where we used a distance modulus
of $m-M = 15.40$ \citep{dur93}.  The right panel shows Ba abundances
as a function of absolute magnitude.  Where available, Ba upper limits
are shown, but stars without Ba measurements are not indicated.  For
reference, the left panel shows Li abundance measurements in both M15
\citep{kir16} and NGC~6397 \citep{lin09}, another metal-poor GC\@.  We
use NGC~6397 as a benchmark because its proximity permits exquisite
spectroscopy even at the main sequence.  The Li measurements in M15
are not of sufficient quality to fully illustrate the mixing episodes.
The detections of Li trace the upper envelope of the true underlying
Li abundance distribution.

The Ba abundances exhibit the $\sim 1$~dex scatter that was already
known to exist in M15.  However, they show no trend with stellar
luminosity.  Excluding upper limits, the mean [Ba/H] abundance at
luminosities fainter than that expected for the FDU is $\bahlo \pm
\bahloerr$, where the mean is weighted by the inverse square of the
uncertainties, and the error bar is the standard error of the mean.
The mean abundance is $\bahhi \pm \bahhierr$ for stars brighter than
the RGB bump, and $\bahmed \pm \bahmederr$ for stars between the two
mixing episodes.  The abundances do not show the expected $\sim
1.5$~dex decrease at the FDU\@.  As expected in any scenario, they
also do not show a decline at the RGB bump.

The fact that the mean abundance is higher for main sequence stars is
almost certainly a result of selection bias.  The Ba measurements
become less certain at fainter magnitudes, as reflected by the
increasing error bars as a function of increasing magnitude in
Figure~\ref{fig:cmdbali}.  Furthermore, about half of the Ba
abundances below the FDU are upper limits.  Therefore, the mean
abundance quoted above should also be viewed as an upper limit.  We
conclude that bias against Ba-poor faint stars causes at least some of
the apparent dependence of Ba abundance on luminosity.  Regardless,
any observed decrease in abundance at the FDU is short of what we
expected in the external pollution scenario.


\section{Discussion}
\label{sec:discussion}

The giants would have lower Ba abundances than main sequence stars in
M15 if Ba and other $r$-process elements originated in a NSM that
occurred long after the cluster finished forming.  We did not observe
the expected decline in abundance.  Therefore, Ba in M15 stars is
well-mixed throughout the star.  We conclude that the $r$-process
elements in M15 were generated before or during the formation of the
cluster. If the enrichment happened before the formation of the
cluster, an inhomogeneity of a factor of 30 in $r$-process abundance
needs to persist over the length scale of the proto-giant molecular
cloud that formed M15. However, such large abundance fluctuations
might be difficult to preserve at parsec scales \citep{mon16}.

This is not the first study to measure the abundances of
neutron-capture elements on the main sequence in globular clusters.
Other clusters with measurements of neutron-capture abundances on or
near the main sequence include M92 \citep{kin98}, M5 \citep{ram03},
47~Tuc, NGC~6397, NGC~6752 \citep{jam04a,jam04b}, and M13
\citep{coh05a}.  In no cluster does the abundance of heavy elements
depend on evolutionary state.  The unique aspect of this study is that
M15 was an especially interesting candidate due to its large scatter
in neutron-capture abundances, which might be expected if the NSM had
preferentially polluted the stars closest to it.

Our study rules out a source with a delay time greater than the
cluster formation time, but it does not necessarily rule out a NSM
altogether.  As mentioned earlier, dense cluster environments can
accelerate the dynamical evolution of compact binaries so that the NSM
delay times would be shorter than in the field \citep{ram15}.
However, M15 is likely not dense enough to cause dynamically driven
NSMs.  Still, \citet{zev19} argued that it is possible for a binary
neutron star system, formed from the first generation of stars in the
GC, to merge within the cluster's star formation timescale
(30--50~Myr).  This scenario requires many conditions to be true,
including Case~BB mass transfer (i.e., after the end of He core
burning) and the ability for the NSM to enrich the cluster despite any
natal kick.  Alternatively, \citet{bek17} proposed that the formation
of M15 might have had a duration of $\sim 0.1$~Gyr, long enough for a
NSM to enrich the stars forming in the cluster.  In support of this
scenario, \citet{ben19} found that about 40\% observed binary neutron
stars in the MW have short ($< 1$~Gyr) merging times.  These include
two systems discovered decades ago with merging times of 0.3--0.4~Gyr
\citep[e.g.,][]{phi91} and three others discovered more recently with
merging times less than 0.1~Gyr \citep[e.g.,][]{sto18}.

In summary, we have ruled out one scenario for $r$-process enrichment
in M15, but the possibilities remain numerous.

\acknowledgments We are grateful to Marc Kassis, Carlos {\' A}lvarez,
and Percy G{\' o}mez for their essential roles in providing DEIMOS
with the 1200B diffraction grating.  We thank Paz Beniamini, E. Sterl
Phinney, and Shrinivas Kulkarni for directing us to the observational
studies of merging times for binary neutron stars.  We also thank Dan
Kasen and Ryan Foley for helpful discussions and the anonymous referee
for a helpful report.

This material is based upon work supported by the National Science
Foundation under Grant No.\ AST-1847909.  E.N.K.\ gratefully
acknowledges support from a Cottrell Scholar award administered by the
Research Corporation for Science Advancement as well as funding from
generous donors to the California Institute of Technology. E.R.-R. and
P.M. thank the Heising--Simons Foundation, the Danish National
Research Foundation (DNRF132), and NSF (AST-1911206 and AST-1852393)
for support.

We are grateful to the many people who have worked to make the Keck
Telescope and its instruments a reality and to operate and maintain
the Keck Observatory.  The authors wish to extend special thanks to
those of Hawaiian ancestry on whose sacred mountain we are privileged
to be guests.  Without their generous hospitality, none of the
observations presented herein would have been possible.

\facility{Keck:II (DEIMOS)}
\software{spec2d \citep{coo12,new13}, MOOG \citep{sne12}, MESA \citep{pax11}}

\startlongtable
\begin{deluxetable*}{lcccccccc}
\tablecaption{Barium Abundances in M15\label{tab:ba}}
\tablehead{\colhead{Star} & \colhead{RA} & \colhead{Dec} & \colhead{$M_{V,0}$} & \colhead{$(B-V)_0$} & \colhead{$v_{\rm helio}$} & \colhead{$T_{\rm eff}$} & \colhead{$\log g$} & \colhead{[Ba/H]} \\
\colhead{ } & \colhead{(J2000)} & \colhead{(J2000)} & \colhead{(mag)} & \colhead{(mag)} & \colhead{(km~s$^{-1}$)} & \colhead{(K)} & \colhead{[cm~s$^{-2}$]} & \colhead{ }}
\startdata
41376 & 21 29 59.34 & $+$12 09 11.9 & $-2.87$ & \phs$ 1.26$ & $ -98.9 \pm  2.1$ & 4231 & 0.68 & $-2.42 \pm 0.12$ \\
40809 & 21 29 59.17 & $+$12 10 16.0 & $-2.56$ & \phs$ 1.15$ & $-109.1 \pm  2.1$ & 4790 & 1.04 & $-2.14 \pm 0.13$ \\
36569 & 21 29 57.94 & $+$12 10 17.0 & $-2.44$ & \phs$ 1.02$ & $-119.5 \pm  2.1$ & 4409 & 0.86 & $-1.95 \pm 0.13$ \\
31227 & 21 29 56.32 & $+$12 09 54.8 & $-2.24$ & \phs$ 1.06$ & $ -97.9 \pm  2.1$ & 4470 & 1.06 & $-2.18 \pm 0.12$ \\
38742 & 21 29 58.55 & $+$12 09 49.8 & $-2.13$ & \phs$ 1.21$ & $-105.1 \pm  2.1$ & 5176 & 1.10 & $-2.98 \pm 0.13$ \\
36913 & 21 29 58.04 & $+$12 09 58.3 & $-1.85$ & \phs$ 0.77$ & $ -88.1 \pm  2.1$ & 4750 & 1.50 &          \nodata \\
41670 & 21 29 59.43 & $+$12 10 11.7 & $-1.85$ & \phs$ 0.91$ & $-103.8 \pm  2.1$ & 4832 & 1.40 & $-2.19 \pm 0.13$ \\
60808 & 21 30 15.67 & $+$12 08 23.3 & $-1.62$ & \phs$ 0.90$ & $-110.2 \pm  2.1$ & 4722 & 1.46 & $-2.32 \pm 0.13$ \\
43593 & 21 30 00.05 & $+$12 09 18.6 & $-1.49$ & \phs$ 0.80$ & $-111.6 \pm  2.1$ & 4958 & 1.61 & $-1.76 \pm 0.13$ \\
54055 & 21 30 06.97 & $+$12 07 46.8 & $-1.48$ & \phs$ 0.72$ & $-112.7 \pm  2.1$ & 5159 & 1.64 &          \nodata \\
16135 & 21 29 43.55 & $+$12 10 03.7 & $-1.33$ & \phs$ 0.79$ & $-101.3 \pm  2.1$ & 4986 & 1.63 & $-1.69 \pm 0.12$ \\
33889 & 21 29 57.17 & $+$12 09 42.6 & $-1.20$ & \phs$ 0.86$ & $-128.9 \pm  2.1$ & 4820 & 1.72 & $-2.25 \pm 0.12$ \\
59959 & 21 30 14.28 & $+$12 09 23.8 & $-1.12$ & \phs$ 0.82$ & $-105.5 \pm  2.1$ & 4771 & 1.70 & $-2.16 \pm 0.13$ \\
48120 & 21 30 01.83 & $+$12 09 49.2 & $-1.06$ & \phs$ 0.75$ & $-113.8 \pm  2.1$ & 4741 & 1.65 & $-2.13 \pm 0.13$ \\
44027 & 21 30 00.19 & $+$12 09 56.2 & $-1.02$ & \phs$ 0.80$ & $-110.7 \pm  2.1$ & 4963 & 1.77 & $-2.47 \pm 0.12$ \\
55914 & 21 30 08.96 & $+$12 08 49.3 & $-1.00$ & \phs$ 0.79$ & $-118.4 \pm  2.1$ & 4850 & 1.76 & $-2.16 \pm 0.13$ \\
49483 & 21 30 02.77 & $+$12 10 38.4 & $-0.97$ & \phs$ 0.66$ & $-115.0 \pm  2.1$ & 5217 & 1.93 & $-1.74 \pm 0.13$ \\
45688 & 21 30 00.76 & $+$12 09 51.9 & $-0.96$ & \phs$ 0.79$ & $ -94.6 \pm  2.1$ & 4901 & 1.78 & $-2.26 \pm 0.12$ \\
51057 & 21 30 04.02 & $+$12 08 58.1 & $-0.87$ & \phs$ 0.79$ & $-109.5 \pm  2.1$ & 4846 & 1.85 & $-2.42 \pm 0.14$ \\
47982 & 21 30 01.75 & $+$12 10 18.7 & $-0.83$ & \phs$ 0.79$ & $-113.8 \pm  2.1$ & 4957 & 1.89 & $-1.99 \pm 0.13$ \\
49428 & 21 30 02.74 & $+$12 09 00.5 & $-0.80$ & \phs$ 0.73$ & $-114.7 \pm  2.1$ & 4716 & 1.84 & $-2.41 \pm 0.15$ \\
32206 & 21 29 56.66 & $+$12 09 52.0 & $-0.63$ & \phs$ 0.59$ & $ -97.5 \pm  2.1$ & 5541 & 2.08 & $-2.54 \pm 0.12$ \\
54512 & 21 30 07.50 & $+$12 10 11.8 & $-0.56$ & \phs$ 0.72$ & $-111.7 \pm  2.1$ & 4964 & 2.06 & $-2.61 \pm 0.13$ \\
34335 & 21 29 57.30 & $+$12 10 19.4 & $-0.53$ & \phs$ 0.69$ & $-121.2 \pm  2.1$ & 4999 & 2.00 & $-2.29 \pm 0.13$ \\
46494 & 21 30 01.06 & $+$12 09 51.3 & $-0.47$ & \phs$ 0.78$ & $-114.8 \pm  2.1$ & 4938 & 1.98 & $-2.62 \pm 0.12$ \\
29005 & 21 29 55.28 & $+$12 09 43.9 & $-0.45$ & \phs$ 0.73$ & $-103.3 \pm  2.1$ & 4871 & 2.00 & $-2.42 \pm 0.12$ \\
42594 & 21 29 59.73 & $+$12 10 10.1 & $-0.43$ & \phs$ 0.74$ & $-105.5 \pm  2.1$ & 5020 & 2.00 & $-2.29 \pm 0.14$ \\
50124 & 21 30 03.27 & $+$12 10 11.1 & $-0.37$ & \phs$ 0.74$ & $-113.1 \pm  2.1$ & 4982 & 2.10 & $-1.65 \pm 0.13$ \\
57863 & 21 30 11.27 & $+$12 10 15.0 & $-0.33$ & \phs$ 0.83$ & $-100.4 \pm  2.1$ & 4842 & 2.13 & $-1.86 \pm 0.13$ \\
52787 & 21 30 05.58 & $+$12 07 05.5 & $-0.27$ & \phs$ 0.72$ & $-112.3 \pm  2.1$ & 5088 & 2.11 & $-2.37 \pm 0.12$ \\
37854 & 21 29 58.30 & $+$12 09 54.1 & $-0.27$ & \phs$ 0.62$ & $-113.2 \pm  2.1$ & 4963 & 2.09 & $-2.25 \pm 0.14$ \\
28721 & 21 29 55.13 & $+$12 10 22.8 & $+0.11$ & \phs$ 0.27$ & $ -92.1 \pm  2.2$ & 6427 & 2.65 & $-2.86 \pm 0.24$ \\
52771 & 21 30 05.57 & $+$12 08 33.0 & $+0.47$ & \phs$ 0.69$ & $-109.0 \pm  2.1$ & 5197 & 2.48 & $-1.74 \pm 0.13$ \\
55135 & 21 30 08.16 & $+$12 08 54.3 & $+0.66$ & \phs$ 0.62$ & $-106.2 \pm  2.1$ & 5173 & 2.58 & $-2.11 \pm 0.12$ \\
22822 & 21 29 50.80 & $+$12 09 51.9 & $+0.81$ & \phs$ 0.61$ & $-102.7 \pm  2.1$ & 5219 & 2.60 & $-2.38 \pm 0.14$ \\
55541 & 21 30 08.54 & $+$12 08 43.4 & $+1.22$ & \phs$ 0.55$ & $-105.0 \pm  2.1$ & 5391 & 2.88 & $-1.92 \pm 0.14$ \\
54237 & 21 30 07.18 & $+$12 08 38.6 & $+1.33$ & \phs$ 0.59$ & $-115.9 \pm  2.1$ & 5352 & 2.89 & $-2.04 \pm 0.13$ \\
17122 & 21 29 44.89 & $+$12 09 21.1 & $+1.39$ & \phs$ 0.58$ & $-106.6 \pm  2.1$ & 5249 & 2.87 & $-2.71 \pm 0.13$ \\
56979 & 21 30 10.17 & $+$12 09 14.6 & $+1.42$ & \phs$ 0.60$ & $-117.6 \pm  2.1$ & 5336 & 2.94 & $-2.36 \pm 0.13$ \\
61191 & 21 30 16.32 & $+$12 08 05.9 & $+1.44$ & \phs$ 0.58$ & $-116.2 \pm  2.1$ & 5471 & 2.98 & $-2.01 \pm 0.14$ \\
57312 & 21 30 10.57 & $+$12 10 27.3 & $+1.49$ & \phs$ 0.63$ & $-117.5 \pm  2.1$ & 5175 & 3.00 & $-1.97 \pm 0.14$ \\
15681 & 21 29 42.91 & $+$12 10 57.3 & $+1.63$ & \phs$ 0.60$ & $-105.0 \pm  2.1$ & 5275 & 3.02 & $-1.99 \pm 0.13$ \\
56947 & 21 30 10.14 & $+$12 07 12.2 & $+1.82$ & \phs$ 0.59$ & $-108.4 \pm  2.2$ & 5492 & 3.13 & $-2.30 \pm 0.20$ \\
8920  & 21 29 28.21 & $+$12 10 15.6 & $+1.93$ & \phs$ 0.57$ & $-110.7 \pm  2.2$ & 5428 & 3.22 & $-2.60 \pm 0.18$ \\
61776 & 21 30 17.39 & $+$12 08 17.1 & $+2.08$ & \phs$ 0.59$ & $-113.0 \pm  2.5$ & 5524 & 3.23 & $-2.35 \pm 0.16$ \\
50825 & 21 30 03.82 & $+$12 09 58.2 & $+2.15$ & \phs$ 0.65$ & $-106.5 \pm  2.2$ & 5349 & 3.20 & $-2.21 \pm 0.23$ \\
61068 & 21 30 16.10 & $+$12 08 14.6 & $+2.17$ & \phs$ 0.48$ & $-115.9 \pm  2.2$ & 5464 & 3.35 & $-1.68 \pm 0.17$ \\
59374 & 21 30 13.38 & $+$12 08 45.2 & $+2.19$ & \phs$ 0.61$ & $-106.9 \pm  2.2$ & 5400 & 3.32 & $-2.07 \pm 0.15$ \\
56251 & 21 30 09.34 & $+$12 06 43.3 & $+2.19$ & \phs$ 0.56$ & $-115.8 \pm  2.2$ & 5431 & 3.25 & $-2.75 \pm 0.18$ \\
31125 & 21 29 56.28 & $+$12 10 16.8 & $+2.20$ & \phs$ 0.67$ & $-111.1 \pm  2.1$ & 4549 & 3.62 & $-1.56 \pm 0.12$ \\
58363 & 21 30 11.91 & $+$12 07 10.7 & $+2.20$ & \phs$ 0.51$ & $-114.8 \pm  2.2$ & 5562 & 3.35 & $-2.25 \pm 0.16$ \\
11998 & 21 29 36.32 & $+$12 08 23.6 & $+2.32$ & \phs$ 0.56$ & $-113.6 \pm  2.2$ & 5621 & 3.36 & $-2.71 \pm 0.18$ \\
59071 & 21 30 12.95 & $+$12 09 46.5 & $+2.34$ & \phs$ 0.54$ & $-108.0 \pm  2.3$ & 5519 & 3.40 & $-1.64 \pm 0.16$ \\
62219 & 21 30 18.20 & $+$12 07 24.8 & $+2.41$ & \phs$ 0.49$ & $-108.5 \pm  2.3$ & 5711 & 3.43 & $-2.11 \pm 0.22$ \\
6374  & 21 29 17.88 & $+$12 10 39.9 & $+2.77$ & \phs$ 0.50$ & $-103.5 \pm  4.9$ & 6042 & 3.67 & $-2.33 \pm 0.26$ \\
7175  & 21 29 21.57 & $+$12 10 20.8 & $+2.80$ & \phs$ 0.50$ & $-108.7 \pm  2.4$ & 5896 & 3.66 & $-2.08 \pm 0.21$ \\
58890 & 21 30 12.65 & $+$12 06 41.7 & $+2.84$ & \phs$ 0.46$ & $-116.9 \pm  2.4$ & 5813 & 3.62 & $-2.95 \pm 0.45$ \\
23303 & 21 29 51.28 & $+$12 08 50.3 & $+3.00$ & \phs$ 0.44$ & $-102.7 \pm  2.3$ & 6098 & 3.83 & $-2.36 \pm 0.25$ \\
62661 & 21 30 19.09 & $+$12 07 48.3 & $+3.12$ & \phs$ 0.34$ & $-105.1 \pm  2.9$ & 6169 & 3.90 &          \nodata \\
18956 & 21 29 47.04 & $+$12 09 48.6 & $+3.14$ & \phs$ 0.35$ & $-109.7 \pm  2.7$ & 6162 & 3.89 & $-2.51 \pm 0.26$ \\
26974 & 21 29 54.04 & $+$12 10 33.2 & $+3.16$ & \phs$ 0.51$ & $-100.4 \pm  2.9$ & 6101 & 3.91 & $-1.81 \pm 0.30$ \\
57247 & 21 30 10.48 & $+$12 07 09.8 & $+3.17$ & \phs$ 0.42$ & $-119.8 \pm  3.1$ & 6292 & 3.97 &        $< -1.44$ \\
17384 & 21 29 45.18 & $+$12 08 54.6 & $+3.18$ & \phs$ 0.40$ & $-101.5 \pm  2.5$ & 6373 & 3.99 &        $< -0.69$ \\
28844 & 21 29 55.20 & $+$12 10 53.9 & $+3.22$ & \phs$ 0.38$ & $-108.6 \pm  2.9$ & 6439 & 4.06 &        $< -0.42$ \\
18422 & 21 29 46.41 & $+$12 09 46.4 & $+3.25$ & \phs$ 0.34$ & $-103.7 \pm  2.5$ & 6213 & 3.93 & $-2.36 \pm 0.30$ \\
22363 & 21 29 50.36 & $+$12 10 30.8 & $+3.35$ & \phs$ 0.33$ & $-104.6 \pm  8.4$ & 6263 & 4.01 & $-1.72 \pm 0.46$ \\
12573 & 21 29 37.50 & $+$12 08 13.7 & $+3.46$ & \phs$ 0.35$ & $-115.3 \pm  3.7$ & 6258 & 4.13 &          \nodata \\
18685 & 21 29 46.73 & $+$12 10 38.1 & $+3.61$ & \phs$ 0.30$ & $-109.6 \pm  5.5$ & 6173 & 4.07 & $-2.46 \pm 0.37$ \\
16142 & 21 29 43.56 & $+$12 08 47.3 & $+3.69$ & \phs$ 0.33$ & $-106.8 \pm  3.5$ & 6557 & 4.29 &        $< -1.66$ \\
27877 & 21 29 54.65 & $+$12 10 58.7 & $+3.69$ & \phs$ 0.55$ & $-100.5 \pm  4.6$ & 5875 & 4.05 &        $< +0.30$ \\
16177 & 21 29 43.61 & $+$12 09 17.1 & $+3.84$ & \phs$ 0.39$ & $-119.4 \pm  3.7$ & 6093 & 4.17 & $-1.92 \pm 0.36$ \\
8227  & 21 29 25.80 & $+$12 11 45.7 & $+3.94$ & \phs$ 0.39$ & $-113.3 \pm  3.4$ & 6709 & 4.38 & $-1.80 \pm 0.37$ \\
12919 & 21 29 38.18 & $+$12 10 32.8 & $+3.95$ & \phs$ 0.35$ & $-100.3 \pm  4.3$ & 6536 & 4.28 &        $< -1.07$ \\
12699 & 21 29 37.77 & $+$12 09 02.4 & $+3.95$ & \phs$ 0.38$ & $-108.3 \pm  3.5$ & 6791 & 4.38 & $-1.71 \pm 0.38$ \\
22494 & 21 29 50.47 & $+$12 08 59.0 & $+4.01$ & \phs$ 0.19$ & $ -99.3 \pm  5.0$ & 5962 & 4.20 &          \nodata \\
7834  & 21 29 24.29 & $+$12 12 10.4 & $+4.05$ & \phs$ 0.42$ & $-115.3 \pm  7.1$ & 6575 & 4.34 &          \nodata \\
7517  & 21 29 23.04 & $+$12 10 54.4 & $+4.06$ & \phs$ 0.42$ & $-109.1 \pm  6.5$ & 6705 & 4.40 &        $< -1.16$ \\
8868  & 21 29 28.04 & $+$12 11 46.2 & $+4.06$ & \phs$ 0.45$ & $-104.7 \pm  5.0$ & 6663 & 4.43 &          \nodata \\
10721 & 21 29 33.37 & $+$12 08 04.5 & $+4.11$ & \phs$ 0.67$ & $ -97.9 \pm  6.2$ & 5010 & 3.97 &        $< -2.04$ \\
25082 & 21 29 52.72 & $+$12 07 52.8 & $+4.13$ & \phs$ 0.48$ & $-114.6 \pm  3.0$ & 6560 & 4.37 &        $< -1.36$ \\
8917  & 21 29 28.20 & $+$12 08 17.2 & $+4.17$ & \phs$ 0.55$ & $-105.4 \pm  9.2$ & 6368 & 4.42 &        $< -1.30$ \\
9864  & 21 29 31.05 & $+$12 09 01.0 & $+4.19$ & \phs$ 0.41$ & $-113.2 \pm  4.0$ & 6323 & 4.36 &        $< -1.48$ \\
7436  & 21 29 22.71 & $+$12 09 00.6 & $+4.23$ & \phs$ 0.34$ & $-119.0 \pm 29.7$ & 6541 & 4.41 & $-1.78 \pm 0.46$ \\
8447  & 21 29 26.60 & $+$12 11 46.1 & $+4.32$ & \phs$ 0.38$ & $-114.5 \pm  7.0$ & 6496 & 4.41 & $-1.80 \pm 0.45$ \\
24115 & 21 29 51.97 & $+$12 10 13.2 & $+4.32$ & \phs$ 0.54$ & $ -97.4 \pm  5.5$ & 6113 & 4.36 &        $< -1.01$ \\
7357  & 21 29 22.35 & $+$12 10 59.0 & $+4.34$ & \phs$ 0.46$ & $-106.5 \pm 16.7$ & 5825 & 4.30 &          \nodata \\
10442 & 21 29 32.61 & $+$12 08 47.1 & $+4.34$ & \phs$ 0.42$ & $-104.6 \pm  3.9$ & 6495 & 4.50 &          \nodata \\
18985 & 21 29 47.07 & $+$12 10 16.2 & $+4.37$ & \phs$ 0.27$ & $-110.7 \pm  7.2$ & 6120 & 4.35 &          \nodata \\
14221 & 21 29 40.57 & $+$12 09 52.3 & $+4.38$ & \phs$ 0.46$ & $-104.9 \pm  8.9$ & 6490 & 4.44 &        $< -1.18$ \\
25509 & 21 29 53.03 & $+$12 09 59.7 & $+4.48$ & \phs$ 0.58$ & $-100.4 \pm  3.7$ & 6186 & 4.49 &          \nodata \\
11536 & 21 29 35.30 & $+$12 08 17.2 & $+4.57$ & \phs$ 0.47$ & $ -86.5 \pm 11.2$ & 6155 & 4.44 &        $< -0.08$ \\
24624 & 21 29 52.37 & $+$12 08 39.7 & $+4.70$ & \phs$ 0.62$ & $ -86.5 \pm 16.3$ & 4750 & 1.50 &          \nodata \\
15876 & 21 29 43.20 & $+$12 10 25.4 & $+4.79$ & \phs$ 0.46$ & $-100.6 \pm  9.3$ & 6249 & 4.50 &        $< -0.85$ \\
8895  & 21 29 28.14 & $+$12 11 18.0 & $+4.80$ & \phs$ 0.45$ & $-117.2 \pm 24.0$ & 5537 & 4.40 &          \nodata \\
14649 & 21 29 41.31 & $+$12 09 35.4 & $+4.99$ & \phs$ 0.47$ & $-104.2 \pm  6.7$ & 6193 & 4.57 &          \nodata \\
11125 & 21 29 34.34 & $+$12 10 44.6 & $+5.10$ & \phs$ 0.51$ & $ -88.0 \pm 16.4$ & 5145 & 4.34 &          \nodata \\
\enddata
\tablerefs{Photometry from \citet{ste94}.}
\end{deluxetable*}

$\,$

\end{document}